\documentclass[10pt]{iopart}
\usepackage[ansinew]{inputenc}
\usepackage[T1]{fontenc}
\usepackage[english]{babel}
\usepackage{amsfonts}

\usepackage{graphicx}
\usepackage{epsfig}
\usepackage{amssymb}
\newtheorem{Theo}{Theorem}
\begin{document}
\title{Topological defects in spinor condensates}
\author{H Mäkelä\dag, Y Zhang\dag\ddag, K-A Suominen\dag\S}  
\address{\dag\ Department of Physics,
University of Turku,
FIN-20014 Turun yliopisto,
Finland}
\address{\ddag\ Departmentof Physics and Institute of Theoretical
  Physics, Shanxi University, Taiyuan 03006, P. R. China} 
\address{\S\ Helsinki Institute of Physics, PL 64, FIN-00014 Helsingin
  yliopisto, Finland}

\begin{abstract}

We investigate the structure of topological defects 
in the ground states of spinor
Bose-Einstein condensates with spin $F=1$ or $F=2$.
 The type and number of defects
are determined by calculating the first and second homotopy groups of
the order-parameter space.
The order-parameter space is identified with a set of degenerate 
 ground state spinors.
  Because the structure of the ground state
depends on whether or not there is an
external magnetic field applied to the system, defects  are 
sensitive to the magnetic field. We study both
 cases and find that the defects in zero and non-zero field are
  different.

\end{abstract}
\section{Introduction}
Bose-Einstein condensates (BECs) of alkali atoms have an 
internal degree of freedom due to the hyperfine spin of these
atoms. If a BEC is realized in a magnetic  trap this degree of
freedom is frozen and in a mean-field limit the  
 condensate is described by a scalar order-
parameter.
 However, if an optical trap \cite{Stamper98}  is used to confine
condensate atoms, this degree of freedom is liberated and  
has to be taken into account \cite{Ho98,Ohmi98}.
  Condensates with this property  are called spinor or vector condensates. 
In the mean-field theory 
the ground state of a spinor condensate is described by an 
order-parameter 
 $\Psi(\mathbf{r})=\sqrt{n(\mathbf{r})}
\xi(\mathbf{r})$, where $n(\mathbf{r})$ is the density of the condensate and
$\xi(\mathbf{r})$ is a normalized  
 spinor, $\xi^\dag (\mathbf{r})\xi(\mathbf{r}) =1$. In this paper 
 the density $n$ is assumed
 to be constant. Because of the
 vectorial nature of the order-parameter, the  behaviour of spinor 
 condensates is in many ways different from that of scalar
 condensates. One manifestation of this can be seen in the difference 
of defects in 
 scalar and spinor condensates. In the former vortices with integer
 winding numbers can exist. The latter allows for more complex
 defects, which are the topic of this paper. Our study is based on the
 ground states calculated using the mean-field theory and single
 condensate approximation \cite{Ho98,Ciobanu, Ueda,Martikainen}.   
Mean-field theory is widely used in the study of Bose-Einstein
condensates and it is usually assumed to give a good description of the
physical system. However, some results suggest that the actual ground
states of spinor condensates may be different from those obtained
using mean-field theory \cite{Law98,Koashi00}. 
Thus, the results of this paper are valid only as long as mean-field theory 
can be applied.

\section{Characterization of the used techniques }
\subsection{The order-parameter space}

In condensed matter systems the concept of an
 order-parameter is very important \cite{Mermin,Mineev98}.
Order-parameter $f(\mathbf{r})$
 is a continuous mapping from some region of the physical space
into the order-parameter space $M$, which consists of all possible values
of the order-parameter. It is usually possible to associate the
order-parameter space with a group $G$ that acts on that space. If
this action is transitive (i.e
for every $x,x'\in M$ there exists some $g\in G$ for which
$x'=g\cdot x$), we can arbitrarily choose some element
$x_{\textrm{ref}}\in M$ which we call the reference order-parameter. 
Every element $x\in M$ can then be obtained from
$x_{\textrm{ref}}$ by acting on it by a suitable element of the
group $G$. Those elements of $G$ which leave $x_{\textrm{ref}}$
fixed constitute a subgroup $H$ called the
isotropy group. Explicitly $H=\{g\in
G\,|\, g\cdot x_{\textrm{ref}}=x_{\textrm{ref}}\}$. Under some
rather general requirements for $G$ and $M$
 the order-parameter space $M$ can be
identified with the quotient space $G/H$.  When considering 
the defects of spin-$F$ Bose-Einstein condensate
 the order-parameter is the
normalized $(2F+1)$-component spinor
$\xi(\mathbf{r})\in\mathbb{C}^{2F+1}$.
 Because we study what kind of  
defects  can exist in the ground state of the system, the order-parameter
space is the set of spinors that minimize the energy.
In the absence of an external  magnetic field we can
 often choose $G=
 U(1)\times SO(3)$. However, this choice is not always the correct one, as
 there may be order-parameter spaces in which $U(1)\times SO(3)$ does
 not act transitively; see
 below. $U(1)\times SO(3)$ acts on $\mathbb{C}^{2F+1}$  via
equation $\big((c,R),\xi\big)\mapsto cD^{(F)}(R)\xi,$ where $ (c,R)\in
U(1)\times SO(3)$ and  $D^{(F)}$ is the $2F+1$ dimensional
irreducible representation of $SO(3)$. This representation is given by
 the map $R(\alpha,\beta,\gamma)\mapsto
 D^{(S)}(\alpha,\beta,\gamma)$, where $R(\alpha,\beta,\gamma)
=R_z(\alpha)R_y(\beta)R_z(\gamma)\in SO(3)$ is given as a product of
 rotations about $y$- and $z$-axes and
$D^{(S)}(\alpha,\beta,\gamma)= \exp(-i\alpha
F_z)\exp(-i\beta F_y)\exp(-i\gamma F_z)$. Here $F_y$ and $F_z$ are $y$- and
 $z$- components of the
spin matrices corresponding to spin $F$.
 Representation matrices for $F=1$ and $F=2$  are given in the 
appendix.

In the presence of an  external magnetic field the energy of the ground
 state is invariant under gauge transformations and rotations about
 the axis of the magnetic field, so we choose $G=U(1)\times SO(2)$.

\subsection{Homotopy groups}

Homotopy groups of the order-parameter space
describe physical defects \cite{Mermin}.
The $n$th homotopy group $\pi_n(M)$ of the
space $M$ consists of the equivalence classes of continuous maps from
$n$-dimensional sphere $S^n$ to the space $M$. Two maps are equivalent
if they are homotopic to one another. In physics, the first
and second homotopy groups are of special importance.
 The first homotopy group $\pi_1(M)$
describes singular line defects and domain walls, which are
non-singular defects. The second homotopy group $\pi_2(M)$ describes
singular point defects and non-singular line defects.
 Thus, identifying $M$ with
$G/H$, we can learn much from the possible defects in
 a physical system
if we know $\pi_1(G/H)$ and $\pi_2(G/H)$. These can be calculated with
the help of the following theorem.

\begin{table}
\caption{\label{Zero1}The reference spinors and their general
 forms  for $F=1$ spinor condensate when the external magnetic field is zero.}
\begin{indented}
\item[]\begin{tabular}{@{}ccc}
\br
& $\xi_{\textrm{ref}}^T$ & $\xi(\alpha,\beta,\gamma,\theta)^T$\\
\mr 
$f$&$(1,0,0)$ &
$e^{i(\theta-\gamma)}(e^{-i\alpha}\cos^2\frac{\beta}{2},
\frac{1}{\sqrt{2}}\sin\beta,
 e^{i\alpha}\sin^2\frac{\beta}{2})$\\
$af$& $(0,1,0)$ & $e^{i\theta}(-e^{-i\alpha}\frac{1}{\sqrt{2}}\sin\beta,
\cos\beta,
 e^{i\alpha}\frac{1}{\sqrt{2}} \sin\beta) $\\
\br
\end{tabular}
\end{indented}
\end{table}

\begin{Theo}\label{theo} Let  $G$ be a Lie group with 
$\pi_0(G)=\pi_1(G)=\pi_2(G)=0$. Here  $0$ denotes a one element group.
  Let $H\subseteq G$ be a closed subgroup, and $H_0\subseteq H$ the
  connected component of the identity. There are isomorphisms 
\begin{equation}
\pi_1(G/H)\cong H/H_0 
\end{equation}
 and 
\begin{equation}
 \pi_2(G/H)\cong\pi_1(H_0).
\end{equation}
\end{Theo} 
 For a proof, see \cite{Mermin} 
or \cite{Poenaru79}.

We cannot use this theorem if  $G$ is $U(1)\times SO(3)$,
because $\pi_1(U(1)\times SO(3))=\mathbb{Z}\times\mathbb{Z}_2$.
 This problem can be solved by using $\mathbb{R}\times SU(2)$ 
instead of $U(1)\times SO(3)$, since this group fulfils the
requirements of the theorem. The former is a covering group of the
latter, 
the covering projection $P:\mathbb{R}\times SU(2)\rightarrow U(1)\times
SO(3)$ being given by $(x,\mathcal{U}(\alpha,\beta,\gamma))\mapsto
\big(e^{ix},R(\alpha,\beta,\gamma)\big)$, where $x\in\mathbb{R}$ and 
\begin{equation}
\mathcal{U}(\alpha,\beta,\gamma)=
\left(\begin{array}{cc}\cos\frac{\beta}{2}e^{-i(\alpha+\gamma)/2}
  &-\sin\frac{\beta}{2}e^{i(\gamma-\alpha)/2}
\\\sin\frac{\beta}{2}e^{-i(\gamma-\alpha)/2} &
\cos\frac{\beta}{2}e^{i(\alpha+\gamma)/2} \end{array}\right)\in SU(2)
\end{equation}
Every matrix in $SU(2)$ can be written in this form. Sufficient 
intervals for $\alpha,\beta$ and $\gamma$ are $[0,2\pi],[0,\pi]$ and
$[0,4\pi]$, respectively.

\section{Spin $1$}
The ground state structure for $F=1$ condensate was calculated by
Ho \cite{Ho98} and by Ohmi and Machida \cite{Ohmi98}. If the external
magnetic field is non-zero the
spin-dependent part in the energy is 
$\mathcal{E}
 (\xi)=c\langle\mathbf{F}\rangle_\xi^2-p\langle F_z\rangle_\xi
+q\langle F_z^2\rangle_\xi $ \cite{Stenger}. Here the
kinetic energy term is neglected in the Thomas-Fermi approximation and
$\langle\mathbf{F}\rangle_\xi=\xi^\dagger\mathbf{F}\xi$. The constant $c$
depends on scattering lengths and the density $n$, whereas $p$ describes
linear and $q$ quadratic Zeeman interaction with the external magnetic
field. The external field is assumed to be directed along the $z$-axis. 
\subsection{Zero external field}
The equation for energy is obtained by setting $p=q=0$.
Depending on the sign of  $c$ energy is minimized either by
$\langle\mathbf{F}\rangle_\xi^2=1$ or
$\langle\mathbf{F}\rangle_\xi=0$. The former is called the ferromagnetic $(f)$
and the latter the antiferromagnetic $(af)$ phase. The order-parameter spaces
corresponding to these phases are $M^f=\{\xi\in\mathbb{C}^3|
\langle \mathbf{F}\rangle_\xi^2=1,\,\xi^\dag\xi=1\} $ and $M^{af}=
\{\xi\in\mathbb{C}^3| \langle
\mathbf{F}\rangle_\xi=0,\,\xi^\dag\xi=1\} $. It is easy to see that
$U(1)\times SO(3)$ acts transitively on these sets. The
reference order-parameters and general order-parameters obtained
from these by a rotation and gauge transformation are shown in
table \ref{Zero1}.

\subsubsection{Ferromagnetic phase}
 From  table $1$ we see that we do not need the angle $\theta$ because
$\gamma$ can produce all possible gauge transformations. This
means that instead of $\mathbb{R}\times SU(2)$ we can use only $SU(2)$.
To find the elements of the isotropy group $H^f$ we set
 $\xi^{f}(\alpha,\beta,\gamma,0)
=\xi^{f}_{\textrm{ref}}$. This gives
$H^{f}=\{\mathbb{I},-\mathbb{I}\}$. Using theorem \ref{theo} we get 
  $\pi_1(G/H^{f})\cong \{\pm\mathbb{I}\}$ and
$\pi_2(G/H^{f})\cong 0$. The order-parameter
space is $SU(2)/\{\pm \mathbb{I}\}\cong SO(3)$.
 This
order-parameter space has also been encountered in ${}^3
\textrm{He}-A$ \cite{Mineyev78}.

Physically, these results mean that we can have one non-trivial
singular vortex but we cannot have any non-trivial monopoles, 
i.e. singular point defects. The non-trivial vortex is a defect
in which the overall phase of the spinor changes by $2\pi$   
 as the defect line is encircled.

\subsubsection{Antiferromagnetic phase}
Now the isotropy group is $H^{af}=\{\big(
 n2\pi,a(\varphi)\big),
\big(( n+\frac{1}{2})2\pi,g a(\varphi)\big)\,|\,\varphi \in
[0,4\pi],
 n\in \mathbb{Z}\}
 $, where we have defined
$a(\varphi)=\mathcal{U}(\varphi,0,0)$ and $g=
 \mathcal{U}(0,\pi,0)$. The connected component
 of the identity is $H^{af}_0
=\{\big(0,a(\varphi)\big)\,|\,\varphi\in [0,4\pi]\}$. 
We get $\pi_1(G/H^{af}) \cong \{\big(n2\pi ,\mathbb{I}\big)
H^{af}_0, \big((n+\frac{1}{2})2\pi,g\big)
H^{af}_0\,|\,n\in \mathbb{Z}\} $.
 This group
 is isomorphic to $\mathbb{Z}$, the isomorphism being given by
the map
 $\big((2n+j)\pi,g^j\big)H^{af}_0\mapsto 2n+j$,
 where $j=0$ or $1$ and $g^0\equiv\mathbb{I}$.
 Thus  $\pi_1(G/H^{af})\cong
 \mathbb{Z}$.  Previously the order-parameter space and 
 first homotopy group  were 
 concluded to be $U(1)\times S^2$ and $\mathbb{Z}$ \cite{Ho98} or
 $[U(1)\times S^2]/\mathbb{Z}_2$ and 
$\mathbb{Z}\times \mathbb{Z}_2$ \cite{Zhou}. Both of these
 order-parameter spaces are incorrect but the first homotopy group in 
\cite{Ho98} is correct. However, that of reference \cite{Zhou} is
 not correct, since there is no (group) isomorphism between $\mathbb{Z}$
 and $\mathbb{Z}\times \mathbb{Z}_2$.  The isotropy group (in $U(1)\times
SO(3)$) is isomorphic to $O(2)$, but it can not be expressed as a
direct product of a subgroup of $U(1)$ and $SO(3)$. Thus the
 order-parameter space can be written only  as
 $G/H=[U(1)\times
SO(3)]/O(2)_{G+S}$, where $G+S$ means that the isotropy group consists
of gauge transformations performed simultaneously with spin
rotations.

\begin{figure}
\begin{center}
\epsfbox{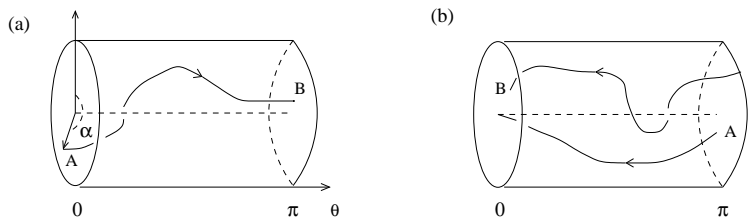}
\end{center}
\caption{\label{Figure}Pictorial representation of the parameters of
    the
antiferromagnetic spinor. Angles $\alpha$ and $\theta$ are shown
in the picture and angle $\beta$ is the length of the vector. On the
    boundary points $(\alpha,\beta,0)$ and $(\alpha\pm
    \pi,\pi-\beta,\pi)$ correspond to the same value of the order-parameter. 
This is also true for all points for which $\beta=\pi$, $\theta$ is fixed
    and $\alpha\in [0,2\pi)$.
A(B) denotes the starting (ending) point of the
curve. We assume $\xi(A)=\xi(B)$, so the curves
define  closed curves in the order-parameter space. Defining the
direction of increasing $\theta$ to be positive, we see in (a) a
path with winding number $\frac{1}{2}$ and in (b) a path  with
winding number $-1$. }
\end{figure}

Because $H^{af}_0$ is homeomorphic
 to $U(1)$ and homeomorphic spaces have the same homotopy groups,
 we get $\pi_2(G/H^{af})\cong
 \pi_2(U(1))\cong\mathbb{Z}$.
If we move around a closed path in the condensate  we note that when
 we  return to the starting point the angle $\theta$ 
 has changed by some amount. If we define the change in
 this angle divided by $2\pi$ to be the winding number, we see
from the elements of $H^{af}/H^{af}_0$
   that the winding number can be either an integer $(n)$ or a half- 
 integer $(\frac{1}{2}+n)$ \cite{Volovik}.
Paths in the order-parameter space can be represented pictorially
as follows. From table 1 we see that we have three parameters in
the general expression for the ferromagnetic state. From these
$\alpha$ and $\beta$ can be restricted to the
intervals $[0,2\pi]$ and $[0,\pi]$, respectively, and
$\theta\in[0,2\pi]$. However, because $\xi^{af}(\alpha\pm
\pi,\pi-\beta,\gamma,\theta)=-\xi^{af}(\alpha,\beta,\gamma,\theta)$,
we can actually restrict $\theta$ to the interval $[0,\pi]$. These
parameters can be represented using cylindrical coordinates
$(\alpha,r,z)$, where now $r=\beta,z=\theta$, see figure \ref{Figure}.

In summary, possible line defects are those in which the overall phase
changes by $2\pi n$ as the defect line is encircled and those in which
a phase change of $\pi+2\pi n$ is accompanied by a $180^\circ $
spinor rotation.
 Also point defects, labelled by integers, are possible.

\begin{table}[h]
\caption{\label{Nonzero1}The ground states
for $F=1$ spinor condensate when the external magnetic field is
non-zero. States are degenerate with respect to angles $\theta$ and $\phi$.
 We have assumed that $c\not =0$. $f_3$ evolves to $f_1$ $(f_2)$
state as $p$ reaches $2c$ $( -2c)$. }
\begin{indented}
\item[]\begin{tabular}{@{}ccc}
\br
& $\xi_{\textrm{ref}}^T$ & $\mathcal{E}(\xi)$\\
\mr
$ f_1$ &$e^{i\theta}(1,0,0)$ & $c-p+q$ \\
$f_2$ & $e^{i\theta}(0,0,1)$& $c+p+q$\\
$f_3$ & $\big(e^{i\theta}\sqrt{\frac{1}{2}+\frac{p}{4c}},0,e^{i\phi}
\sqrt{\frac{1}{2}-\frac{p}{4c}}\big)$ & $-\frac{p^2}{4c} +q$\\
$af$& $e^{i\theta}(0,1,0)$ &$0$ \\
\br
\end{tabular}
\end{indented}
\end{table}

\subsection{Non-zero external field}
Straightforward minimization of energy gives four ground states which
 are degenerate with respect to one or two  phase variables, see
  table \ref{Nonzero1}.  The identification of the
 order-parameter space $G/H$ is easier than in the absence of the
 magnetic field.

 In $f_1,f_2$ and $af$ states
 $G=U(1)$ and $H={1}$, so $G/H= U(1)$ for which
$\pi_1(U(1))\cong\mathbb{Z}$ and $\pi_2(U(1))\cong 0$. Thus we
 can have singular vortices with arbitrary integer winding numbers but we
 do not have singular point defects. This resembles the situation in a 
scalar condensate, where we have similar defects.

In $f_3$ state 
$G/H=U(1)\times U(1)$ for which  $\pi_1(U(1)\times U(1))\cong 
\mathbb{Z} \times\mathbb{Z}$ and $\pi_2(U(1)\times U(1))\cong 0$. Now
 we can have independent  vortices
in the $m=1$ and $m=-1$ components of the spinor.

\section{Spin $2$}

The ground states for $F=2$ spinor condensate were calculated by
Ciobanu \emph{et al} \cite{Ciobanu} and  Ueda and Koashi
\cite{Ueda}.
In the Thomas-Fermi approximation the spin-dependent
 energy  is given by
$\mathcal{E}(\xi)=c\langle \mathbf{F}\rangle_\xi^2
+d|\Theta_\xi|^2-p\langle F_z\rangle_\xi$,
 where $c$ and $d$ are constants depending on
scattering lengths and $p$ describes the linear Zeeman effect.
 The possible ground states are characterized by two
 parameters, namely $||\langle \mathbf{F}\rangle_\xi||
=(\langle\mathbf{F}\rangle_\xi^2)^{1/2}$ and
$|\Theta_\xi|=|2\xi_2\xi_{-2}-2\xi_1\xi_{-1} +\xi_0^2|$.

\subsection{Zero external field}

The energy in zero magnetic field is obtained by setting $p=0$.
Because $\langle\mathbf{F}\rangle_\xi^2$ and
$|\Theta_\xi|$ are invariant under the action of $U(1)\times
SO(3)$, and for each $\xi$ there exists a rotation $R$ for which
$\langle \mathbf{F}\rangle_\xi^2=\langle F_z\rangle_{D(R)\xi}^2$,
we can write the energy in the form
 $\mathcal{E}(\xi)=c\langle F_z\rangle_\xi^ 2
+d|\Theta_\xi|^2$. This equation has been solved in
\cite{Ciobanu,Ueda}. It turns out that there are  three possible phases
in the system, ferromagnetic $(F,F')$, cyclic $(C)$ and polar $(P)$.

To express the order-parameter spaces we  define
$M(i,j)=\{\xi\in \mathbb{C}^5\,|\,
||\langle \mathbf{F}\rangle_\xi||=i,|\theta_\xi|=j,\xi^\dag\xi=1\}$.
Then in $F$ phase the order-parameter space is $M(2,0)$, in  $F'$
phase  $M(1,0)$, in $C$ phase $M(0,0)$ and in $P$ phase
$M(0,1)$. Representative spinors from these sets and their energies
 are shown in table \ref{Zero2}.
It turns out that $U(1)\times SO(3)$ acts
transitively on the parameter spaces of the ferromagnetic and cyclic
phases.
However, this is not true for the order-parameter space of the
polar phase. For example, if we
first choose $\phi=0$ and then $\phi=\frac{\pi}{2},\psi=0$ in the
spinor representing a polar state,  we get two spinors which can not
be converted to each other by a rotation and gauge  transformation.
This means $U(1)\times SO(3)$ is not a group large enough in the case
of polar phase.

\begin{table}
\caption{\label{Zero2} Ground state spinors and their
energies of $F=2$ condensate when the external magnetic field is
absent.
 General forms of the ground states
can be obtained from
  these by a rotation and a gauge transformation. Notice that in $P$
  state we have two free parameters. Those are needed because we
  cannot obtain every possible spinor representing the polar state from a
  fixed reference spinor by a rotation and a gauge transformation. }
\begin{indented}
\item[]\begin{tabular}{@{}cccc}
\br
& $\xi^T$&$\mathcal{E}$ \\
\mr
$F$& $(1,0,0,0,0) $& $4c$    \\
$F'$&
$(0,1,0,0,0)$ & $c$ \\
$C$&
$ \frac{1}{2}( 1,0,\sqrt{2},0,-1)$ &$0$ \\
$P$
&$\frac{1}{\sqrt{2}}\big(\sin\phi\sin\psi,\sin\phi\cos\psi,\sqrt{2}\cos\phi,
-\sin\phi\cos\psi,\sin\phi\sin\psi\big)$& $d$\\
\br
\end{tabular}
\end{indented}
\end{table}

\subsubsection{Ferromagnetic phases}
There are two possible ferromagnetic phases, labelled  by $F$
and $F'$. As in $F=1$ case, in both of these phases 
we can use $SU(2)$ instead of
$\mathbb{R}\times SU(2)$.

In $F$-phase $H^F=H^F/H^F_0=\{\mathbb{I},(-i\sigma_z),(-i\sigma_z)^2,
(-i\sigma_z)^3\}$ and thus 
 $\pi_1(G/H^F)\cong\mathbb{Z}_4$, $\pi_2(G/H^F)\cong 0$. Here $\sigma_z$ is 
the $z$-component of Pauli matrices and $-i\sigma_z$ describes
rotation about the $z$-axis through $180^\circ$. Non-trivial vortices are
those in which the reference spinor rotates through $180^\circ,360^\circ$ or
$540^\circ$ about $z$-axis when the defect line is circulated.
 In $SO(3)$ the isotropy group is $\{\pm \mathbb{I}\}$, 
so $G/H=SO(3)/\mathbb{Z}_2$. For a pictorial representation
of paths in $SO(3)/\mathbb{Z}_2$ see \cite{Vollhardt}. 

In $F'$ phase the order-parameter space is $SO(3)$, and
defects are similar to those 
in the ferromagnetic phase of spin-1 condensate.

\subsubsection{Cyclic phase}

In $C$-phase we meet an example of a non-commuting first homotopy
group. A rotation and a gauge transformation of the reference spinor
give
\begin{displaymath}
\xi^C=\frac{1}{2}e^{i\theta}
\left(\begin{array}{c}
e^{-2i\alpha}\big(\cos^4\frac{\beta}{2}e^{-i2\gamma}
+\frac{\sqrt{3}}{2}\sin^2\beta
-\sin^4\frac{\beta}{2}e^{i2\gamma}\big)\\e^{-i\alpha}\sin\beta
\big(\cos^2\frac{\beta}{2}e^{-i2\gamma}-\frac{\sqrt{3}}{2}\sin 2\beta
+\sin^2\frac{\beta}{2}e^{i2\gamma}\big)\\
-i\frac{\sqrt{6}}{2}\sin^2\beta\sin 2\gamma
+\frac{\sqrt{2}}{4}(1+3\cos2\beta
)\\e^{i\alpha}\sin\beta\big(
\sin^2\frac{\beta}{2}
e^{-i2\gamma}+\frac{\sqrt{3}}{2}\sin 2\beta
+\cos^2\frac{\beta}{2}e^{i2\gamma}\big)
\\e^{2i\alpha}\big(\sin^4\frac{\beta}{2}e^{-i2\gamma}
+\frac{\sqrt{3}}{2}\sin^2\beta
-\cos^4\frac{\beta}{2}e^{i2\gamma}\big)\end{array}\right)
\end{displaymath}
 Equating this with $ \frac{1}{2}\left( 1,0,\sqrt{2},0,-1\right)^T$
 yields the elements of the isotropy group.
$H^C$ turns out to be a discrete, non-commuting group. Explicitly  $H^C$ is
the union of the conjugacy classes shown below.
 The isotropy group (in $U(1)\times SO(3))$ is
  isomorphic to the tetrahedral group, which is 
 the symmetry group of a tetrahedron. 
 $H^C$ is a discrete group and thus
$\pi_1(G/H^C)=H^C$. Because $H^C$ is a non-commuting group we have to use the
conjugacy classes of $\pi_1(G/H^C)$ to classify the topologically
 inequivalent defects \cite{Mermin}. Two line defects
are topologically equivalent if and only if they are characterized
by the same conjugacy class of the first homotopy group. Defects can
 still be labelled by the elements of the first homotopy group, but if
 these elements  belong to the same conjugacy class, corresponding 
defects can be continuously transformed to one another. However, if 
they belong to different conjugacy classes this is not possible. The conjugacy
classes are

\begin{equation}\begin{array}{lll}
\fl C_0(n) = \{(n,\mathbb{I})\},\quad
\overline{C_0}(n)=\{(n,-\mathbb{I})\}\\
\fl C_2(n)=\{(n, a),(n,-a), (n, b),(n,-b),(n, c),(n,-c)\}\\
\fl C_3(\frac{1}{3}+n)=\{
(\frac{1}{3}+n,d),(\frac{1}{3}+n, e),
(\frac{1}{3}+n,f),(\frac{1}{3}+n,g)\}\\
\fl \overline{C_3}(\frac{1}{3}+n)=\{(\frac{1}{3}+n, -d),(\frac{1}{3}+n,-e),
(\frac{1}{3}+n,-f),(\frac{1}{3}+n,-g)\}\\
\fl C_3^2(\frac{2}{3}+n)=\{(\frac{2}{3}+n,d^2),(\frac{2}{3}+n,e^2),
(\frac{2}{3}+n,f^2),(\frac{2}{3}+n,g^2)\}\\
\fl \overline{C^2_3}(\frac{2}{3}+n)=\{(\frac{2}{3}+n,-d^2),
(\frac{2}{3}+n, -e^2),(\frac{2}{3}+n,-f^2),(\frac{2}{3}+n,-g^2)\}
\end{array}\end{equation}
Here 
$n\in\mathbb{Z},\,a=\mathcal{U}(\pi,0,0),b=\mathcal{U}(0,\pi,\frac{\pi}{2}),
c=\mathcal{U}(0,\pi,\frac{3\pi}{2}),
d=\mathcal{U}(\frac{\pi}{4},\frac{\pi}{2},\frac{\pi}{4}),
e=\mathcal{U}(\frac{\pi}{4},\frac{\pi}{2},\frac{13\pi}{4}),
f=\mathcal{U}(\frac{13\pi}{4},\frac{\pi}{2},\frac{\pi}{4}),
g=\mathcal{U}(\frac{5\pi}{4},\frac{\pi}{2},\frac{13\pi}{4})$
 and $a^2=b^2=c^2=d^3=e^3=f^3=g^3=-\mathbb{I}$.
 We have also  divided the real number part of each group element 
 by $2\pi$. The
class $C_0(n)$ describes defects in which the phase of the spinor
is changed by $2\pi n$ as the defect line is encircled. Notice that only 
 $C_0(0)$ corresponds to trivial defects. In the
case of $\overline{C_0}(n)$ phase change of $2\pi n$ is accompanied by 
 a $360^\circ$ rotation about $z$-axis. 
For the rest of the conjugacy classes an explicit description of the
defects is  more complicated.
For example the element $(n,a)$ in the class
$C_2(n)$ depicts a defect in which the  spinor rotates through
$180^\circ$ about  $z$-axis and changes phase by $2\pi n$ as the
line is encircled. Similarly $(n,b)\in C_2(n)$ describes rotations
first through $90^\circ$ about the $z$-axis and then through
$180^\circ$ about the  $y$-axis together with $2\pi n$ phase
change. However, because these defects belong to the same conjugacy
class they can be continuously transformed into one other.

The
multiplication table of conjugacy classes is shown in table \ref{Conjugacy}.
 It  shows that, for example, when we combine defect $C_2(n)$
with  $C_2(-n)$ they can either annihilate each other $(C_0(0))$ or form
defect $\overline{C_0}(0)$ or $C_2(0)$, the result depending on how
they are brought together.  

Defects can be classified further using homology groups 
\cite{Kleman77,Trebin82}. 
In the presence of other line singularities it may be possible 
to transform two line defects described by different conjugacy classes
 into one another. This is achieved 
 by splitting a defect into two parts and 
combining these  beyond a suitable line defect. Elements of $\pi_1(M)$
can be grouped into sets in such a way that defects described by
elements in the same set can be deformed to one another either
continuously or by the previously described way. 
The collection of  these sets forms a factor group  
 $\pi_1(M)/D$, where $D$ is an  invariant subgroup of $\pi_1(M)$
 generated by elements $\delta\tau\delta^{-1}\tau^{-1}$ with $
\delta,\tau\in \pi_1(M)$. The elements of $\pi_1(M)/D$ are unions of
conjugacy classes. In our case  $D$ is the union of the conjugacy classes with
 winding number zero, 
$D=C_0(0)\cup\overline{C}_0(0)\cup
C_2(0)$, and  
\begin{equation}\label{D}
\pi_1(G/H^C)/D=\{C_0\cup \overline{C}_0\cup C_2
,C_3\cup \overline{C_3} ,C_3^2\cup 
\overline{C_3^2}\}.
\end{equation}
 Here we have omitted winding numbers,
which are $n,1/3+n$ and $2/3 +n$ respectively.
 We see that line defects with the same winding number can 
be deformed to one another either continuously or 
using a splitting and recombination process.

From the work of Poenaru and Toulouse \cite{Poenaru} we know that when two
line defects (described by $\delta,\tau\in\pi_1(M)$) cross each
other they produce a new line defect connecting them. This defect is
of the type $\delta\tau\delta^{-1} \tau^{-1}$. Clearly, if 
$\delta\tau\delta^{-1}\tau^{1}=1$, line defects can pass through each
other without the creation of a new singular defect. In our case
defects that can be created by making two line defects  cross 
 are the trivial defect  $C_0(0)$ and two
non-trivial defects, namely $\overline{C_0}(0)$ and $C_2(0)$.

\begin{table}
\caption{\label{Conjugacy} The multiplication table of the
conjugacy classes of $C$-phase. Because the class multiplication
is commutative only half of that is shown. Winding numbers have been
omitted for clarity. When two classes
are multiplied the winding number of the resulting class is the
sum of the individal winding numbers. }
\begin{indented}
\item[]
\begin{tiny}
\begin{tabular}{@{}l|llllllllll}
\br
&$\overline{C_0}$ &$C_2$ &$ C_3$ &$\overline{C_3}$
&$C_3^2$ &$\overline{C_3^2}$\\
\mr
$\overline{C}_0$ &$C_0$ &&&&&\\
$C_2$ &$C_2$ &$6C_0+6\overline{C_0}+4C_2$ \\
$C_3$ &$\overline{C_3}$ &$3(C_3+\overline{C_3})$  
&$3\overline{C_3^2}+C_3^2$ &&&& \\
 $\overline{C_3}$ &$C_3$ &$3(C_3+\overline{C_3})$ &$\overline{C_3^2}+3C_3^2$
 &$3\overline{C_3^2} +C_3^2$&&& \\
 $C_3^2$ &$\overline{C_3^2}$ &$3(C_3^2+\overline{C_3^2})$ 
&$4\overline{C_0}+2C_2$
 &$4C_0+2C_2 $ &$3C_3+\overline{C_3}$ &&\\
 $\overline{C_3^2}$ &$C_3^2$ &$3(C_3^2+\overline{C_3^2})$ &$ 4C_0+2C_2$
&$4\overline{C_0}+2C_2$
&$C_3+3\overline{C_3}$ &$3C_3+\overline{C_3}$\\
\br
\end{tabular}
\end{tiny}
\end{indented}
\end{table}

\subsection{Non-zero external field}
Ground states were calculated in \cite{Ciobanu,Ueda} and
 are shown in table \ref{Nonzero2}. However, now it should be noted that in the
 cyclic phase the order-parameter space has a quite complicated structure 
\cite{Ueda}. Group $U(1)\times SO(2)$ can act transitively on this order
 parameter space only if the external field is strong enough, and even
 then there may be states which are degenerate in energy but which can
 not be obtained from the reference order-parameter shown in 
table \ref{Nonzero2} \cite{Ueda}.

 In the ferromagnetic phases and in the $P_0$
phase $G/H=U(1)$ and the first and second homotopy groups are
$\mathbb{Z}$ and $0$.
In $C,P$ and $P_1$ phases $G/H=U(1)\times U(1)$ and the homotopy
groups are $\mathbb{Z}\times\mathbb{Z}$ and $0$. Physically this
means that we can have a vortex in each component of a spinor but only
two of them can have independent winding numbers.
In $C$-phase, if there are vortices with winding
numbers $m$ and $n$ say, in the first and third component of the
spinor, then there must be a vortex also in the fifth component of the 
spinor. However, its winding number is not free but equal to $2n-m$.

\begin{table}
\caption{\label{Nonzero2} General forms of  the  ground state
    spinors of $F=2$ condensate  in the precence of external magnetic
    field. Energy is degenerate with respect to
 angles $\theta$ and $\phi$.}
\begin{indented}
\item[]
\begin{tabular}{@{}cccc}
\br
& $\xi^T$&$\mathcal{E}$ \\
\mr
$F_1$& $e^{i\theta}(1,0,0,0,0) $& 4c-2p    \\
$F_2$& $e^{i\theta}(0,0,0,0,1) $& 4c+2p    \\
$F'_1$&$e^{i\theta}(0,1,0,0,0)$ &c-p \\
$F'_2$&$e^{i\theta}(0,0,0,1,0)$ &c+p \\
$C$&$ \frac{1}{2}\big(
e^{i\theta}(1+\frac{p}{4c}),0,e^{i\phi}\sqrt{2-\frac{p^2}{8c^2}
},0,e^{-i(\theta-2\phi)}(-1+\frac{p}{4c})\big)$& $-\frac{p^2}{4c}$ \\
$P$&$\frac{1}{\sqrt{2}}\big(e^{i\theta}\sqrt{1+\frac{p}{4c-d}}
,0,0,0,e^{i\phi}\sqrt{1-\frac{p}{4c-d}}\big)$& $d-\frac{p^2}{4c-d}$\\
$P_1$ &$\frac{1}{\sqrt{2}}\big(0,e^{i\theta}\sqrt{1+\frac{p}{2(c-d)}},0,
e^{i\phi}\sqrt{1-\frac{p}{2(c-d)}},0\big)$& $d-\frac{p^2}{4(c-d)}$\\
$P_0$ &$e^{i\theta}(0,0,1,0,0)$ &$d$\\
\br
\end{tabular}
\end{indented}
\end{table}

\section{Discussions}
In this paper, we have calculated  the first and second homotopy groups
of the order-parameter spaces of spinor condensates with $F=1$ and
$F=2$. 
The elements of these groups 
correspond to topologically stable singular line and point defects.
The order-parameter space is identified with the set of degenerate
ground state spinors, and both non-zero and zero external magnetic field cases
 are discussed. 

In $F=1$ condensate 
there are two possible phases, ferromagnetic and
antiferromagnetic. 
If external field is zero in the former there can be one topologically
non-trivial line defect but no topologically non-trivial 
 point defects. In the latter 
 infinitely many line and point defects, labelled by integers, are
 possible. 

In $F=2$ condensate three different phases, ferromagnetic, polar and
cyclic are possible. The ferromagnetic phase can be further divided
into two phases labelled by 
$||\langle \mathbf{F}\rangle_\xi||=1$ or $2$. In zero
field the former has similar defects to the ferromagnetic phase of $F=1$ 
condensate and  
in the latter there can be three topologically non-trivial line 
defects but point defects are not stable.   

 In the absence of an external field the order-parameter
 space of the cyclic phase has a
non-commuting first homotopy group. Topologically stable defects are
 classified by the conjugacy classes of this group and are 
those in which  the spinor is suitably rotated and its phase changed by an
integer multiple of $\pi/3$ as the defect line is encircled. Stable 
 point defects are not possible. If
 external magnetic field is applied the symmetry is reduced and 
non-commutativity of the first homotopy group is lost. It also turns out 
that in the zero field  $U(1)\times SO(3)$ does not act transitively on the
order-parameter space of the polar phase 
 and thus the  defect structure remains unsolved.

For $F=1$ and $F=2$ condensates, if the external 
field is non-zero and there is only one
non-zero component in the spinor, a vortex with an  
arbitrary integer winding number is possible. 
If there are two or three non-zero 
components then a vortex in each component of the spinor is possible, but
only two of these can have an independent winding number. 
In the presence of a magnetic field stable point
defects cannot exist. 
    
\ack
We thank the Academy of Finland (project 50314) for financial
support. YZ is also supported by NSF of China (grant nos 10175039 and 
90203007).

\appendix
\section*{Appendix}
\setcounter{section}{1}
The $3\times 3$ representation matrix corresponding to $\mathcal{U}
(\alpha,\beta,\gamma)$ is
\begin{equation}\fl  D^{(1)}(\alpha,\beta,\gamma)=
\left(\begin{array}{ccc}
e^{-i(\alpha+\gamma)}\cos^2\frac{\beta}{2}
 &-e^{-i\alpha}\frac{1}{\sqrt{2}}\sin\beta
 &e^{-i(\alpha-\gamma)}\sin^2\frac{\beta}{2}\\
e^{-i\gamma}\frac{1}{\sqrt{2}}\sin\beta&\cos\beta & -e^{i\gamma}
\frac{1}{\sqrt{2}}\sin\beta\\
 e^{i(\alpha-\gamma)}\sin^2\frac{\beta}{2}&
 e^{i\alpha}\frac{1}{\sqrt{2}} \sin\beta
& e^{i(\alpha+\gamma)}\cos^2\frac{\beta}{2}
\end{array}\right)
 \end{equation}
The five dimensional representation matrix is given by
 $D^{(2)}(\alpha,\beta,\gamma)=\exp(-i\alpha F_z)\exp(-i\beta
  F_y)\exp(-i\gamma F_z)$, where $\exp(-i\alpha
  F_z)=diag(e^{-i2\alpha},e^{-i\alpha},1,e^{i\alpha},e^{i2\alpha})$
and

\begin{footnotesize} \begin{equation}\begin{array}{ll}
&\fl \exp(-i\beta F_y)\\&\fl =\left(\begin{array}{ccccc}
\cos^4\frac{\beta}{2}&-
\sin\beta\cos^2\frac{\beta}{2}& \frac{\sqrt{6}}{4}
\sin^2\beta
&-\sin\beta\sin^2\frac{\beta}{2}&
\sin^4\frac{\beta}{2}\\
\sin\beta\cos^2\frac{\beta}{2}&\frac{1}{2}
(\cos\beta +\cos
2\beta)& -\frac{\sqrt{6}}{4}\sin 2\beta &
\frac{1}{2}(\cos\beta-\cos
2\beta)&-
\sin\beta\sin^2\frac{\beta}{2}\\ \frac{\sqrt{6}}{4}
\sin^2\beta&
 \frac{\sqrt{6}}{4}\sin 2\beta &\frac{1}{4}(1+3\cos 2\beta)
&-\frac{\sqrt{6}}{4}\sin 2\beta
 &\frac{\sqrt{6}}{4}\sin^2\beta \\
 \sin\beta\sin^2\frac{\beta}{2}
 & \frac{1}{2}(\cos \beta -\cos 2\beta)&
\frac{\sqrt{6}}{4}\sin 2\beta
&\frac{1}{2} (\cos\beta +\cos
2\beta)&-\sin\beta\cos^2\frac{\beta}{2}
\\\sin^4\frac{\beta}{2}&
\sin\beta\sin^2\frac{\beta}{2} &\frac{\sqrt{6}}{4}
\sin^2\beta&
\sin\beta\cos^2\frac{\beta}{2}
& \cos^4\frac{\beta}{2}
\end{array}\right)\end{array} \end{equation}
\end{footnotesize}

\end{document}